# Suppression of Intertwined Density Waves in $La_4Ni_{3-x}Cu_xO_{10+\delta}$

Stephen Zhang[1,*], Danrui Ni[1], Ruyi Ke[1], Guangming Cheng[2], Nan Yao[2], Robert J. Cava[1,†]

*sz3870@princeton.edu

† rcava@princeton.edu

1 Department of Chemistry, Princeton University, Princeton New Jersey 08544, USA.

2 Princeton Materials Institute, Princeton University, Princeton New Jersey 08544, USA

## Abstract

Superconductivity in $La_4Ni_3O_{10}$ has been reported to emerge upon suppression of intertwined spin and charge density wave (SDW/CDW) order, suggesting a possible connection to the pairing mechanism. Here we report a systematic investigation of $La_4Ni_{3-x}Cu_xO_{10+\delta}$ ($0 \leq x \leq 0.7$), focusing on the evolution of the SDW/CDW order as a function of chemical substitution. Temperature-dependent resistivity, magnetic susceptibility, and Hall effect measurements reveal a linear suppression of density wave transition temperature $T_{dw}$ and a concurrent enhancement of hole concentration with increasing Cu content. At higher substitution levels ($x > 0.15$), the transition induced anomaly in the resistivity becomes undetectable while a magnetic signature persists, indicating a partial decoupling of spin and charge components and the possible survival of short-range spin correlations. The absence of superconductivity across the substitution series highlights the importance of additional factors in stabilizing the superconducting state in pressurized $La_4Ni_3O_{10}$.

## Introduction

The Ruddlesden-Popper (RP) nickelates ($A_{n+1}B_nO_{3n+1}$, where n ≠ 0, ∞) have drawn considerable interest due to their complex electronic, magnetic, and structural behavior, particularly after the reports of high-pressure superconductivity in $La_3Ni_2O_7$ and $La_4Ni_3O_{10}$ [1–15]. In $La_4Ni_3O_{10}$ layered nickelate, the existence of intertwined spin-density wave (SDW) and charge-density wave (CDW) orders have been evidenced by kinks in resistivity and magnetic susceptibility measurements, as well as incommensurate charge/spin superlattice structures using single crystal x-ray and neutron diffraction [16-17]. A similar coupling of charge and spin modulations has been observed in cuprates and other layered nickelates, where the periodicity of the antiferromagnetic spin order is exactly half that of the charge modulation [18-23]. In $La_4Ni_3O_{10}$, these density wave phases compete with the emergence of superconductivity under large hydrostatic pressure and are reported to be completely suppressed in the superconducting state. A structural phase transition from monoclinic P21/a to tetragonal I4/mmm occurs right before the emergence of the superconducting phase, with the SDW being a competing order potentially linked to the pairing mechanism [2,11].

Although high-pressure studies have played a central role in uncovering superconductivity in bulk Ruddlesden-Popper nickelates, more recent work has shown that epitaxial strain—arising from substrate mismatch—can also stabilize superconductivity in $La_3Ni_2O_7$ thin films, albeit with a reduced $T_C$ [13,24]. This builds on the earlier reports of superconductivity in nickelates, which was first observed in infinite-layer $Nd_{0.8}Sr_{0.2}NiO_2$ thin films [25–30]. Further investigation into the interplay between electronic phases, lattice dynamics, and competing orders in these Ruddlesden-Popper nickelates is essential to elucidate the mechanisms underpinning both their superconductivity and density wave behavior. While thin film studies of $La_4Ni_3O_{10}$ have yet to report superconductivity or density wave modulation, and pressure remains to date the only demonstrated method of tuning its electronic phases, we propose here a different strategy. Chemical substitution offers an alternative pathway to manipulate these competing orders. In this work, we investigate the effect of Cu substitution on the suppression of density wave states in $La_4Ni_{3-x}Cu_xO_{10+\delta}$. By partially substituting Ni with Cu, we observe the suppression of the density wave $T_{dw}$ as well as the magnitude of the resistive signature. The hole-type carrier concentration increases linearly with Cu content, and plateaus once the density wave is fully suppressed. We observe an increase in carrier density that appears to be caused by the delocalization of charge from the suppression of the density wave order. These results demonstrate that chemical substitution effectively decouples the spin and charge components of the density wave order and drives significant carrier delocalization, offering a new handle for tuning the correlated electronic phases in trilayer nickelates.

**Results**

Temperature-dependent resistivity measurements of $La_4Ni_{3-x}Cu_xO_{10+\delta}$ for $0 \leq x \leq 0.27$ are presented in **Fig. 1**. The density wave transition temperature, $T_{dw}$, is defined by the minimum in the first derivative of resistivity [16]. For the parent compound $La_4Ni_3O_{10+\delta}$, shown in **Fig. 1a**, we observed a pronounced kink at $T_{dw} \approx 132K$, indicating the onset of the SDW/CDW transition. Below this temperature, resistivity slightly increases, consistent with reduced carrier mobility due to the formation of a charge/spin superlattice. A low-temperature upturn in resistivity is also observed, likely resulting from scattering at polycrystalline grain boundaries, as such features are absent in single-crystal measurements of $La_4Ni_3O_{10+\delta}$ [31]. As Cu content increases, $T_{dw}$ systematically shifts to lower temperatures. The magnitude of the transition, reflected in both the kink in $\rho(T)$ and the inverted peak in $\frac{d\rho}{dT}$, diminishes with increasing Cu substitution and becomes nearly undetectable for $x \geq 0.15$. For comparison, previous studies on electron-doped $La_4Ni_{3-x}Al_xO_{10}$ revealed the opposite trend at low substitution levels, where both the transition temperature and magnitude were enhanced [32].

Ambient pressure and temperature scanning transmission electron microscopy of the $La_4Ni_3O_{10}$ structure is presented in **Fig. 2a**. The triple stacking of the perovskite structure can be seen with the out-of-plane vector parallel to the monoclinic b-axis. The first derivative of resistivity with respect to temperature, shown in **Fig. 2b**, highlights the progressive suppression of both the transition temperature and its magnitude as Cu content increases. **Figure 2c** displays the hole concentration determined by Hall measurements (depicted in **Fig. S1**, see the Supplemental Material [33]) as a function of Cu substitution, revealing a linear increase in carrier density with x that plateaus for $x \geq 0.3$. Thermogravimetric analysis (TGA), shown in **Fig. S2** for $x = 0$, was used to determine the oxygen content in $La_4Ni_{3-x}Cu_xO_{10+\delta}$, with all samples annealed under high oxygen pressure exhibiting excess oxygen in **Fig. 2d**. A gradual increase in δ with increasing Cu content was observed, suggesting that interstitial oxygen may contribute to the rising hole concentration. However, the delocalization of charge carriers due to suppression of spin- and charge-density wave (SDW/CDW) order cannot be excluded as an additional factor contributing to the enhanced carrier density.

Temperature-dependent magnetic susceptibility measurements were carried out for $La_4Ni_{3-x}Cu_xO_{10+\delta}$ and are shown in **Fig. 3**. For the undoped $La_4Ni_3O_{10+\delta}$, a subtle kink in $\chi(T)$ is observed near the same temperature as the resistive anomaly, consistent with the onset of the spin-density wave order. As Cu substitution increases, the anomaly gradually shifts to lower temperatures, mirroring the trend observed in the resistivity measurements, and is indicative of a progressive suppression of magnetic order. However, even at higher Cu concentration $x \geq 0.15$, the anomaly in $\chi(T)$ remains discernible as its resistive counterpart becomes weak or absent. For $x > 0.3$, the anomaly disappears, such that no kink can be observed in the susceptibility. The evolution of $\chi(T)$ thus corroborates the resistivity and supports the interpretation of substitution induced suppression of the SDW order. **Figure 4** summarizes the evolution of the density wave transition temperature, $T_{dw}$, with respect to Cu content in $La_4Ni_{3-x}Cu_xO_{10+\delta}$, based on resistivity and magnetic susceptibility measurements, and includes carrier concentration with respect to $x$. The phase diagram reveals a near linear suppression of $T_{dw}$ with increasing Cu substitution, with transition temperature values in agreement for both transport and magnetic measurements, confirming that both anomalies originate from a common spin-density wave transition. The onset of the plateau in the carrier concentration coincides with the complete suppression of the density wave transition at $x \approx 0.3$.

**Discussion**

The suppression of the density wave transition in $La_4Ni_{3-x}Cu_xO_{10+\delta}$ proceeds linearly with Cu substitution, at a rate of approximately -13.8 K per 1% Cu substitution. This behavior is similar to the effects observed in undoped $La_4Ni_3O_{10}$ under applied pressure, where the transition is suppressed at a comparable rate of -13K/GPa and vanishes prior to the onset of superconductivity [34]. Hall effect measurements reveal a significant increase in hole concentration with increasing Cu substitution. While TGA indicates a gradual rise in oxygen nonstoichiometry δ with increasing Cu substitution, the carrier density gain far exceeds what would be expected from oxygen uptake alone. Assuming full ionization of excess oxygen, the contribution to hole concentration from δ would yield an increase on the order of $10^{20}$ cm$^{-3}$ across the series; however, the measured increase in carrier density reaches $10^{22}$ cm$^{-3}$.

One potential explanation is that the suppression of the SDW/CDW order releases a significant fraction of carriers localized due to strong correlations or trapped in charge ordered states. Similar behavior has been documented in underdoped cuprates, where the nominal hole doping level often overestimates the number of mobile carriers. Hall effect and optical conductivity measurements have consistently shown that the number of mobile carriers is often less than the number of doped holes in stripe-ordered or pseudogapped phases [19,35]. However, when spin and charge density wave orders are suppressed—via temperature, pressure, magnetic field, or further doping—a sudden increase in carrier mobility and concentration is observed. Notably, in $YBa_2Cu_3O_{6+x}$, Badoux et al. reported a jump in the Hall number from $p$ to approximately $p+1$ near the pseudogap critical point, indicating that the suppression of charge order delocalizes a significant number of previously immobile holes [36]. Similarly, Cu doping of $TiSe_2$ enhances carrier concentration while suppressing charge density wave as a precursor to the superconducting phase [37]. These findings underscore that the emergence of enhanced metallicity, and ultimately superconductivity, can stem not only from doped carriers but also from the release of carriers bound in ordered states. This behavior may explain the observations made in $La_4Ni_{3-x}Cu_xO_{10+\delta}$, where the increase in carrier concentration significantly exceeds what can be explained by oxygen stoichiometry alone, suggesting that the suppression of density wave order plays a central role in carrier delocalization. At high Cu substitution levels ($x \geq 0.3$), the carrier concentration begins to plateau, coinciding with the complete suppression of the density wave transition from the disappearance of the magnetic susceptibility anomaly. This suggests that once the spin/charge ordering is fully suppressed, further Cu substitution does not contribute additional mobile carriers. The delocalization of previously localized carriers, associated with the breakdown of SDW/CDW order, reaches its limit in this regime.

A notable divergence occurs at higher Cu concentrations ($x > 0.15$), where the resistive anomaly associated with the density wave transition is almost completely suppressed, while the kink remains visible in the magnetic susceptibility. This separation suggests a partial decoupling of the spin and charge components of the density wave order at higher Cu concentration. This divergence in CDW and SDW components has been similarly observed in cuprates, for example, in $La_{1.48}Nd_{0.4}Sr_{0.12}CuO_4$, where it was shown that charge density wave sets in a higher temperature than the spin order [19, 38]. These results support the interpretation that the spin-density wave in $La_4Ni_{3-x}Cu_xO_{10+\delta}$ may evolve into a robust, disordered or short-range state as the long-range charge order is suppressed, with the magnetic signature remaining detectable. The signature of the charge ordering may be too weak to observe in resistivity measurements, and further temperature dependent TEM studies may depict this ordering in samples with greater Cu-content. At higher Cu-substitution, no superconductivity was observed in $La_4Ni_{3-x}Cu_xO_{10+\delta}$ for $x \leq 0.7$. For $x \geq 0.7$, $La_2NiO_4$ began to form as a secondary phase as seen in **Fig. S3** via powder x-ray diffraction, with increasing concentration for greater Cu substitution. Supplementary **Fig. S4** shows the energy dispersive x-ray spectroscopy data for $La_4Ni_{2.3}Cu_{0.7}O_{10+\delta}$, where the addition of the 214 nickelate based on the relative stoichiometry can be seen. This discrepancy between $La_4Ni_{3-x}Cu_xO_{10+\delta}$ and undoped $La_4Ni_3O_{10}$ under pressure underscores the likely importance of the structural phase transition, monoclinic P2$_1$/a to tetragonal I4/mmm, in stabilizing the superconducting phase, a transformation that is absent in the Cu-substituted series.

**Conclusion**

In summary, we have demonstrated that Cu substitution in La$_4$Ni$_{3-x}$Cu$_x$O$_{10+\delta}$ systematically suppresses the density wave transition, as evidenced by the progressive weakening and eventual disappearance of resistivity and magnetic susceptibility anomalies associated with the spin-density wave order. The transition temperature $T_{dw}$ decreases linearly with increasing Cu content, and no signature of a structural phase transition or superconductivity emerges within the substitution range studied. These results underscore the sensitivity of the SDW phase to chemical substitution and suggest that both the suppression of competing orders and a structural transformation may be necessary for superconductivity to emerge in La$_4$Ni$_3$O$_{10}$. It remains unclear whether the suppression of the SDW/CDW is driven primarily by increased hole concentration or by disorder introduced through random copper substitution, ongoing neutron scattering and cryogenic transmission electron microscopy (cryo-TEM) experiments will offer direct insight into the evolution of spin and charge ordering, potentially helping to resolve these questions. Future studies combining chemical substitution and pressure could decouple the roles of carrier density and lattice symmetry.

**Methods and Materials**

Polycrystalline La$_4$Ni$_{3-x}$Cu$_x$O$_{10-\delta}$ was synthesized using the sol-gel method. The starting materials La$_2$O$_3$ (Sigma-Aldrich 99.9%), Cu(NO$_3$)$_2$ · 6H$_2$O (Sigma-Aldrich 99.9%), and Ni(NO$_3$)$_2$ · 6H$_2$O (Sigma-Aldrich 99.9%) were measured in appropriate ratios and dissolved in approximately 4M nitric acid with a small amount of citric acid. The mixture was stirred and heated in a 95°C water bath until a green gel formed. The resulting gels were heated at 250°C until a fluffy brown powder formed, which was ground and reheated to 800°C to burn off any remaining organic components. The resulting black powder was pressed into pellets and heated to 1200°C under 4 bar of O$_2$ for a duration of 24 h. To obtain the correct oxygen stoichiometry, the pellets were annealed at 1100°C under 36 bar of O$_2$ for 24 h and subsequently furnace cooled.

Powder x-ray diffraction (PXRD) was carried out using a Bruker D8 Advance Eco diffractometer with LYNXEYE detector and Cu Kα radiation. Pieces of La$_{4-x}$Ni$_{3-x}$Cu$_x$O$_{10-\delta}$ were ground for several minutes in an agate mortar and pestle to form a fine powder. Hall effect measurements were taken from -3 to +3 T at 1.8 K using a Quantum Design Physical Properties Measurement System (PPMS). Positive-field resistance values were subtracted from negative-field resistance values to remove quadratic terms from the data. DC resistivity measurements between 1.8K and 300K were taken using the Quantum Design resistivity module. Electrical contacts were made using silver epoxy and cured at 120°C. DC magnetic measurements were taken using the Quantum Design vibrating sample magnetometer module. Thermogravimetric analysis was performed using a TA Instruments TGA 5500, with 5% H$_2$ forming gas flow at 2C/min ramping rate to 800C. Scanning electron microscopy and energy dispersive x-ray spectroscopy was performed using a Quanta 200 FEG Environmental-SEM

Thin lamellae were prepared by focus ion beam cutting using a Helios NanoLab G3 UC dual-beam focused ion beam and scanning electron microscope (FIB/SEM) system. Sample thinning was accomplished by gently polishing the sample using a 2 kV Ga+ ion beam in order to minimize surface damage caused by the ion beam. Conventional transmission electron microscope (TEM) imaging and atomic resolution high-angle annular dark-field (HAADF) STEM imaging were performed on a Titan Cubed Themis 300 double Cs-corrected scanning/transmission electron microscope (S/TEM), operated at 300 kV.

**Acknowledgement**

The initial work was performed with the support of the US Department of Energy grant DE-FG02-98ER45706, but after its suspension, by Princeton University, and the Imaging and Analysis Center is partly supported by the PCCM MRSEC and partly by Princeton University. The authors acknowledge the use of Princeton's Imaging and Analysis Center, which is partially supported by the Princeton Center for Complex Materials, a National Science Foundation (NSF)- MRSEC program (DMR-2011750). S.Z. synthesized samples, performed x-ray diffraction, electrical and magnetic measurements, and conducted data analysis. D.N. performed thermogravimetric analysis. R.K. performed scanning electron microscopy. G.C. and N.Y. performed transmission electron microscopy. R.C. assisted with data analysis and conceptualization. S.Z. wrote the manuscript. Authors declare that they have no competing interests. All data required to evaluate the conclusions in the paper are present in the paper and/or Supplementary Materials. Additional data related to this paper may be requested from the authors.

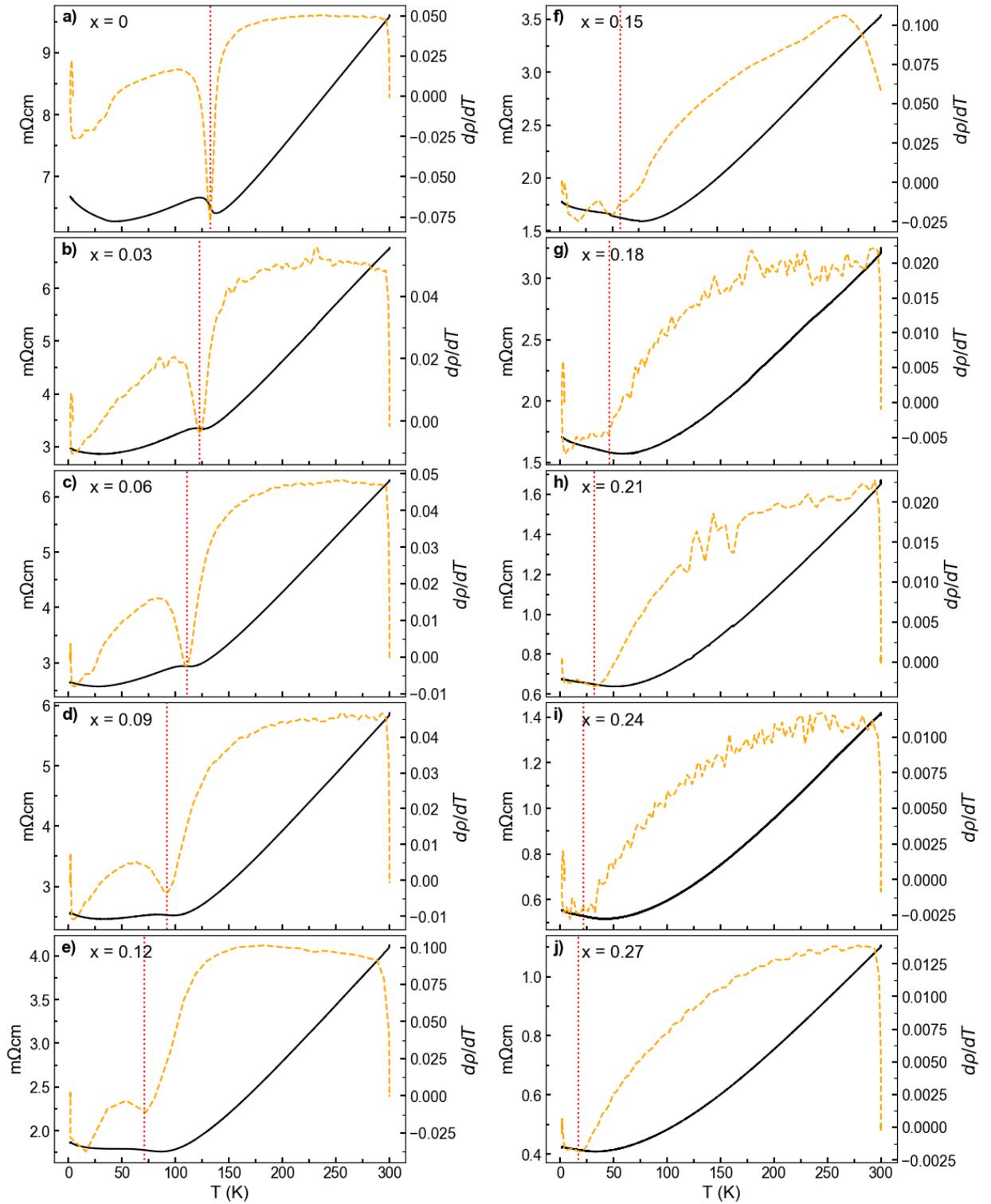

**Figure 1**. Temperature dependent resistivity of $La_4Ni_{3-x}Cu_xO_{10+\delta}$ for selected doping levels, measured from 0 to 300K. The black solid line depicts resistivity, while the orange dashed lines show the first derivative in resistivity with respect to temperature. The density wave transition, $T_{dw}$, is marked at the minima of $\frac{d\rho}{dT}$ using a vertical red dotted line. As Cu content increases, $T_{dw}$ shifts to lower temperature and the magnitude of the resistive anomaly progressively weakens. The width of the transition broadens with Cu substitution as well. At approximately $x = 0.15$, the inverted peak seen previously in the first derivative is nearly fully suppressed.

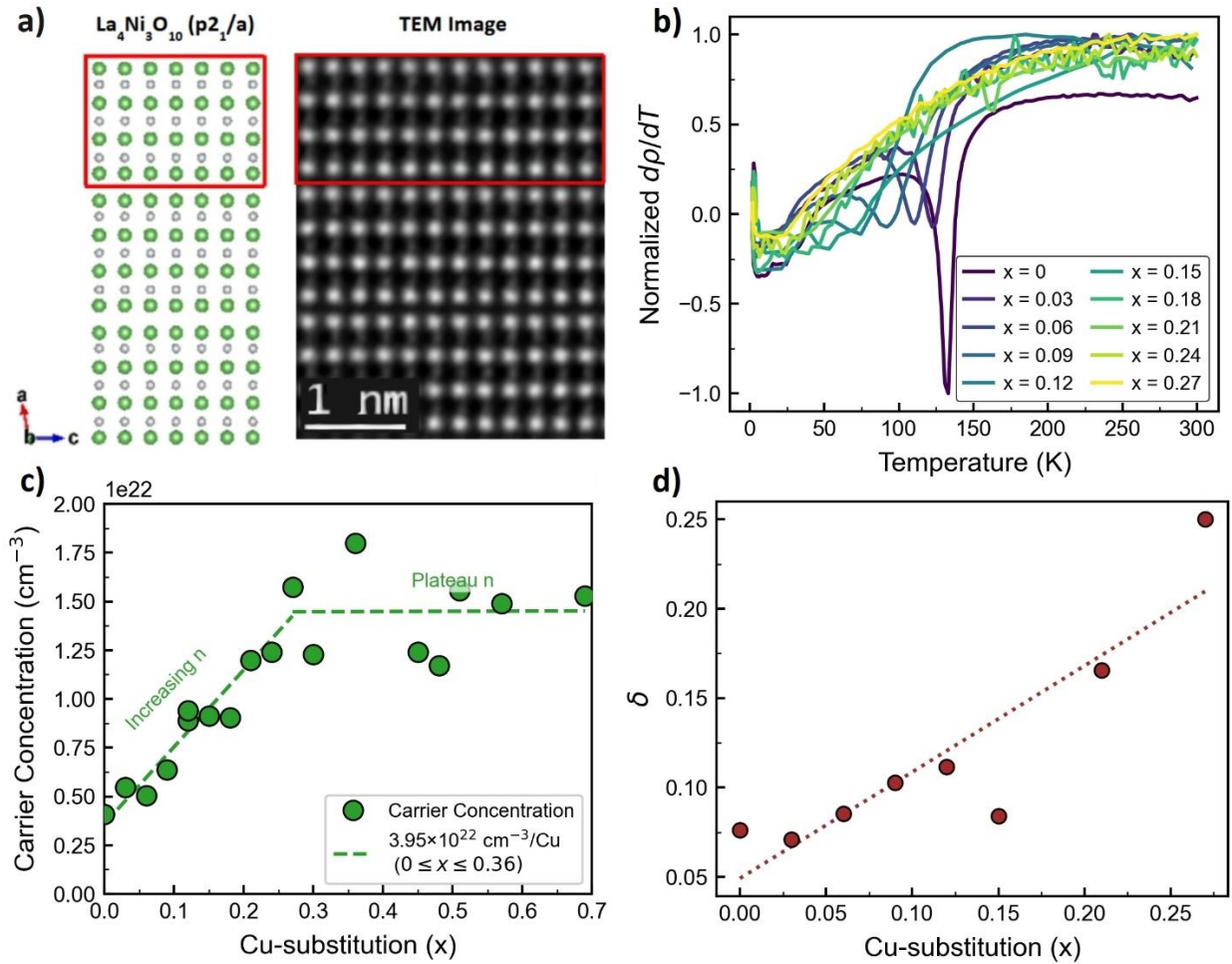

**Figure 2**. **a)** A high angle annular dark field (HAADF) scanning transmission electron microscopy image of an undoped $La_4Ni_3O_{10+\delta}$ polycrystalline sample is shown alongside the monoclinic phase of $La_4Ni_3O_{10}$ in the ac-plane. The first derivative of resistivity with respect to temperature $\frac{d\rho}{dT}$ is compiled in **b)**, highlighting the suppression of the density wave transition with increasing Cu substitution. **c)** Carrier concentration (*n*), determined from Hall effect measurements at 1.8 K, increases linearly with Cu substitution. A linear fit was performed for $0 \leq x \leq 0.36$, where *n*

plateaus for greater *x*. **d)** Oxygen non-stoichiometry ($\delta$), derived from thermogravimetric analysis, shows a gradual increase in oxygen content with high Cu substitution levels.

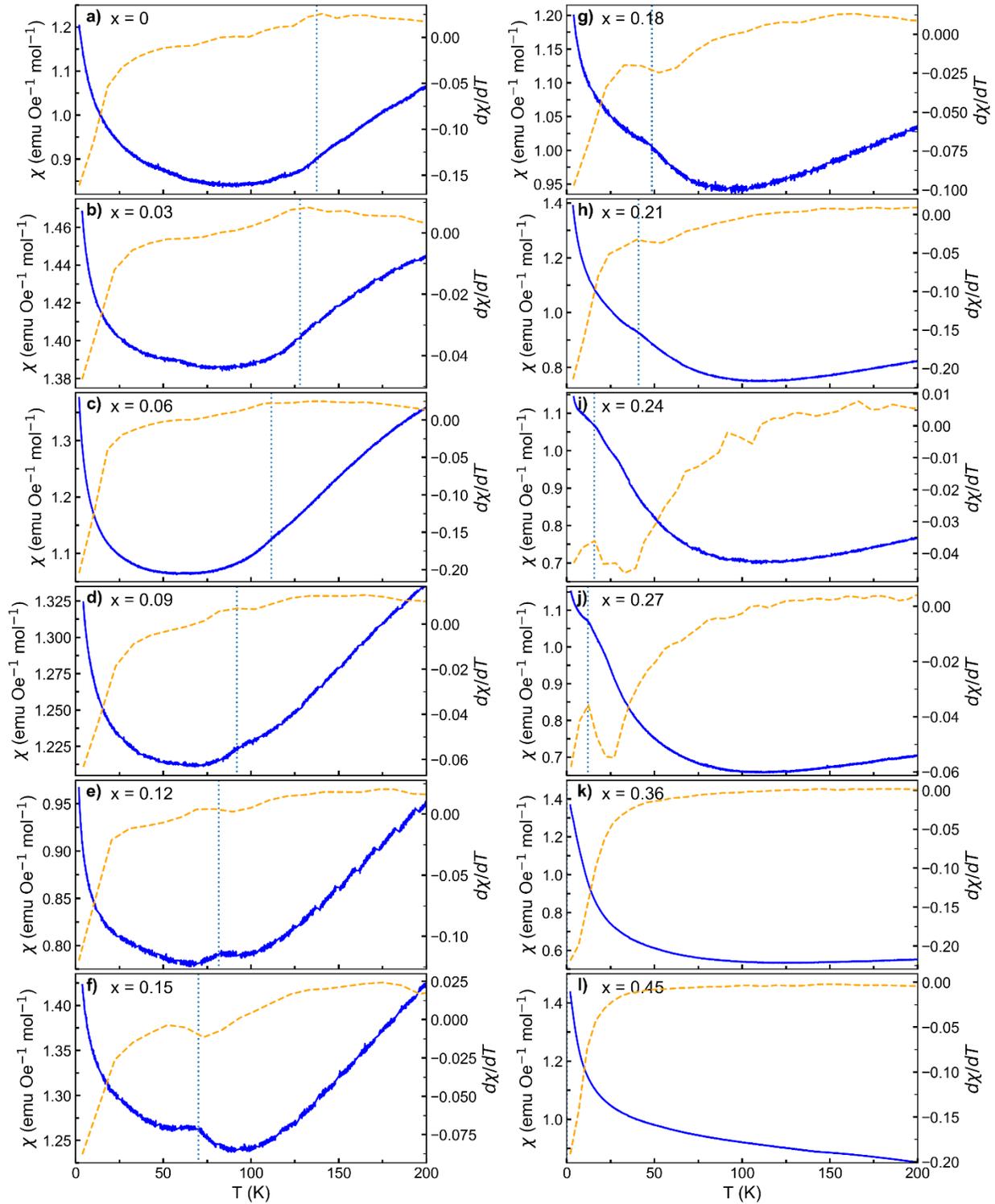

**Figure 3**. Temperature-dependent magnetic susceptibility ($\chi$) for $La_4Ni_{3-x}Cu_xO_{10+\delta}$ is shown in solid blue lines. The orange dashed lines represents the temperature-dependent first derivative of the magnetic susceptibility. The density wave transition, extracted from the first derivative, is indicated by the vertical blue dotted lines. The suppression of the transition with increasing Cu content closely mirrors the trend observed in resistivity measurements, however, the transition signature persists for $x > 0.15$, and the width of the transition doesn't broaden with Cu substitution, unlike in the resistivity. For $x > 0.3$, the transition is fully suppressed and no kink is observed in the magnetic susceptibility, as $\chi$ becomes largely paramagnetic.

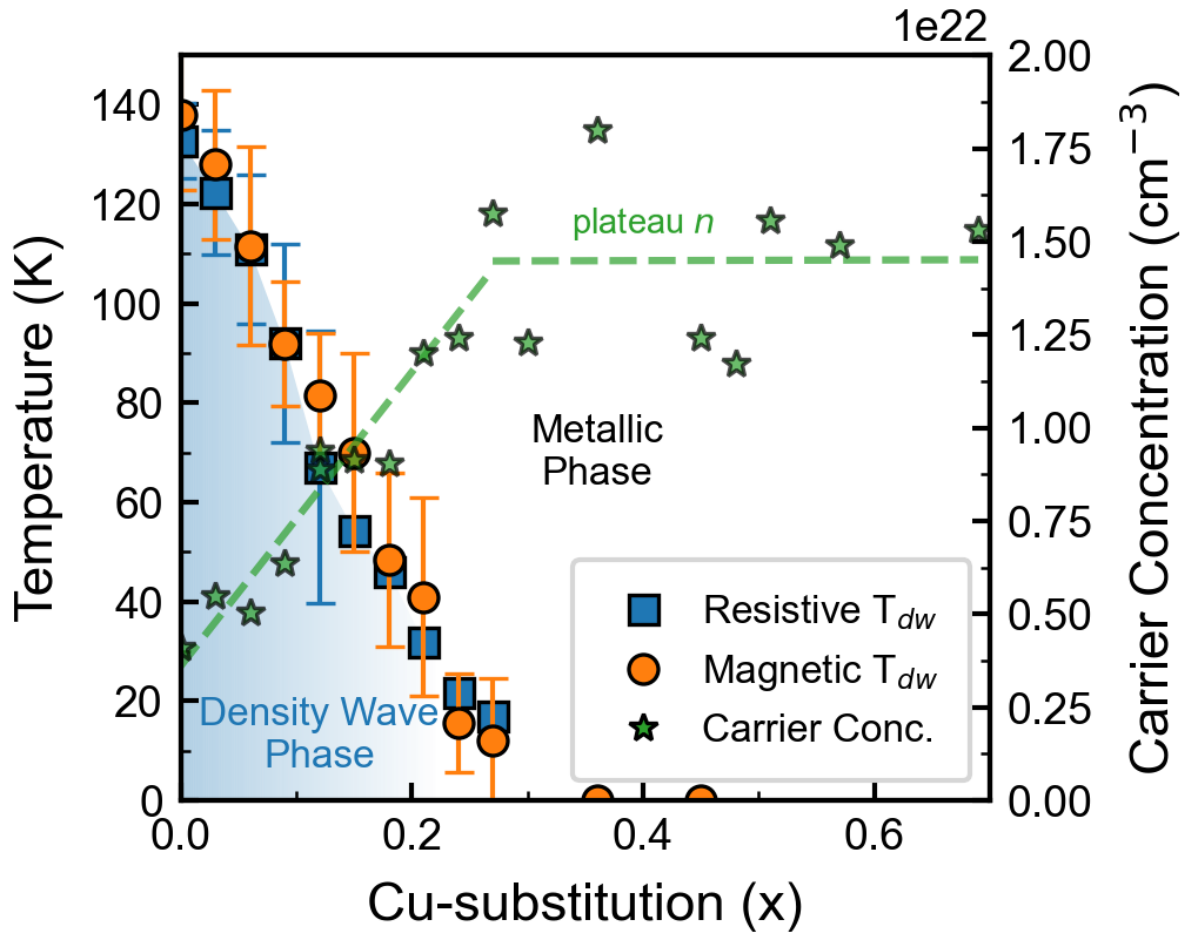

**Figure 4**. Phase diagram of La$_4$Ni$_{3-x}$Cu$_x$O$_{10+\delta}$ showing the evolution of the density wave transition temperature (T$_{dw}$) as a function of Cu content. Blue square markers correspond to values extracted from resistivity data, and orange circle markers from magnetic susceptibility. The resistive transition temperature decreases linearly with Cu substitution and vanishes near x ≈ 0.20, indicating the complete suppression of long-range spin-density wave order, whereas the magnetic transition isn't fully suppressed until $x \approx 0.3$. The linear fit shown has a slope of -13.8K per 1% Cu-substitution, with $R^2 = 0.98$. The error bars were taken using the full-width half maximum/minimum of the peak seen in $\frac{d\rho}{dT}$ and $\frac{d\chi}{dT}$. The green star markers represent the carrier concentration, where it increases for $x < 0.3$ and subsequently plateaus for $x > 0.3$, which matches the location of the full suppression of density wave phase in the magnetic susceptibility, or when $T_{dw}$ approaches zero.